\documentclass{aa}  

\usepackage{graphicx}
\graphicspath{{figures/}}
\usepackage[varg]{txfonts}
\usepackage{commath}
\usepackage[super]{nth}

\usepackage{siunitx}
\DeclareSIUnit{\erg}{erg}
\DeclareSIUnit{\jansky}{Jy}
\DeclareSIUnit{\parsec}{pc}

\usepackage[dvipsnames]{xcolor}
\definecolor{click_color}{RGB}{46,48,118}

\usepackage{hyperref}
\hypersetup{
  colorlinks=true,
  linkcolor=click_color,
  citecolor=click_color,
  urlcolor=click_color
  }
\usepackage[capitalize]{cleveref}
\crefname{equation}{eq.}{eqs.}
\crefname{section}{Sect.}{Sects.}

\begin{document} 

\title{FRB~20121102A monitoring: Updated periodicity in the L band.}

\author{
    C. A. Braga\thanks{Email: cristobal.braga@ug.uchile.cl}\inst{1,2}
    \and
     M. Cruces\inst{3,4,5,6,2}
     \and
     T. Cassanelli\inst{7}
     \and
    M.C. Espinoza-Dupouy\inst{1}
    \and
   L. Rodriguez \inst{8}
    \and
    L. G. Spitler \inst{6}
    \and
    J. Vera-Casanova \inst{2,5}
    \and
    P. Limaye \inst{6,9}
    }

\institute{
    Departament of Astronomy, Universidad de Chile, Camino El Observatorio 1515, Las Condes, Santiago, Chile
    \and
    Centre of Astro-Engineering, Pontificia Universidad Católica de Chile, Av. Vicuña Mackenna 4860, Santiago, Chile
    \and
    Joint ALMA Observatory, Alonso de Córdova 3107, Vitacura, Santiago, Chile
    \and
    European Southern Observatory, Alonso de Córdova 3107, Vitacura, Casilla 19001, Santiago de Chile, Chile
    \and
    Department of Electrical Engineering, Pontificia Universidad Católica de Chile, Av. Vicuña Mackenna 4860, Santiago, Chile
    \and
    Max-Planck-Institut für Radioastronomie, Auf dem Hügel 69, D-53121 Bonn, Germany
    \and 
    Department of Electrical Engineering, Universidad de Chile, Av.~Tupper 2007, Santiago 8370451, Chile
    \and
     Instituto de Astrofísica, Facultad de Física, Pontificia Universidad Católica de Chile, Casilla 306, Santiago 22, Chile
     \and
     Argelander Institute for Astronomy, 53121 Bonn
    }
              
\date{Received September XX, XXXX; accepted March XX, XXXX}


\abstract
{FRB~20121102A was the first fast radio burst to be observed to repeat. Since then, thousands of bursts have been detected by multiple radio telescopes around the world. Previous work has shown an indication of a cyclic activity level with a periodicity of around 160 days. Knowing when the source repeats is essential for planning multi-wavelength monitoring to constrain the emission extent and progenitor source.}
{We report the monitoring of FRB~20121102A using the 100-m Effelsberg radio telescope in the L band and update the periodicity of the cyclic activity level.}
{We used the Lomb-Scargle periodogram on a sample of 284 observing epochs, of which\SI{42}{\percent} correspond to detections and \SI{58}{\percent} to non-detections. Our dataset is composed of the seven epochs of our monitoring plus publicly available data. We investigated two methods: i) a binary model, describing the observing epochs with \num{1} if there are detections and with 0 for non-detections, and ii) a normalised rates model that considers the inferred detection rates.}
{We report no detections in 12.5-hour observations down to a fluence of \SI{0.29}{\jansky\milli\s}. The best period we find for the cyclic activity window is \num{159.3 +- 0.8} days for the binary model and \num{159.3 +- 0.2} days for the normalised rates model. We show the activity phase to be \SI{53}{\percent}. The normalised rates show clear Gaussian-like behaviour for the activity level, in that the number of detections peaks at the centre of the activity window.}
{The periodicity found through both methods is consistent for the L- and S-band datasets, implying it is intrinsic to the source. The activity phase in the S band however,  shows an indication of it ending before the L-band activity phase, supporting the idea of a chromatic dependence of the activity window. The sample in the C band is not large enough to further confirm this result.}


\maketitle

\section{Introduction}
\label{sec:introduction}

Fast radio bursts (FRBs) are highly energetic (\SIrange{e36}{e41}{\erg}) radio pulses, with durations ranging from microseconds to milliseconds, which come from extragalactic sources \citep{2007Sci...318..777L}. They are characterised by a frequency-dependent time delay in their arrival times, which is quantified by the dispersion measure (DM; in units of \si{\parsec\per\centi\m\cubed}) and corresponds to the column density of free electrons between the observer and the source. Based on their detection, we can categorise them into two types: one-offs, which have been detected only once, and repeaters. Morphological studies of FRBs show that one-offs and repeaters have distinguished spectral and time structures, \citep{2021ApJ...923....1P} suggesting that they come from different populations and potentially have different origins. Of the repeaters only two sources exhibit periodic 
activity windows: FRB~20180916B, \citep{2020Natur.582..351C} with a periodicity of \num{16.35} days, and FRB~20121102A, with a periodicity of \num{157} days found in \cite{2020MNRAS.495.3551R} and \num{161.3} days found later in \cite{2021MNRAS.500..448C}. FRB~20180916B has an active window (the phase where the bursts are detected) of \SI{\sim31}{\percent} which means that the source is active for 5 days out of the 16 \citep{2020Natur.582..351C}. For FRB~20121102A the activity window is \SI{\sim60}{\percent}, which means that out of the 161 days the source is active for roughly \num{97} days \citep{2021MNRAS.500..448C}. The first repeater ever detected, FRB~20121102A \citep{2016Natur.531..202S}, was localised by very long baseline interferometric observations to its host, a low-metallicity, star-forming dwarf galaxy at redshift $z = \num[separate-uncertainty=false]{0.19273(8)}$ \citep{2017Natur.541...58C,2017ApJ...834L...7T}. One of the explanations for the periodic behaviour of FRBs is a formation scenario involving a binary system, where the periodicity arises from the orbital period. Another possibility, pointed out for FRB~20180916B, is that the periodicity is due to the precession of a magnetar \citep{2024MNRAS.530.3641F}. Even though we still do not know the origin of these FRBs, their periodic behaviour makes follow-up observations and multi-wavelength campaigns easier, allowing us to characterise their emission extent and therefore constrain their progenitor source.

In this paper, we report the follow-up observations of 
FRB~20121102A in the L band using the 100-m Effelsberg radio 
telescope
and the updated periodicity results. In 
\cref{sec:observation_details}, we describe the setup of our 
observations. In \cref{sec:data_processing}, we present the 
techniques used to combine our dataset with all publicly 
available observations of the source. In 
\cref{sec:results_and_periodicity_calculation}, we present the
results of the follow-up and the updated periodicity and 
activity window. In \cref{sec:discussion}, we discuss the 
implications of periodicity for FRB~20121102A, and in 
\cref{sec:conclusions}, we present our final remarks and 
conclusions.

\begin{table}[t]
	\centering
	\caption{FRB~20121102A monitoring setup with Effelsberg.}
	\label{tab:example_table}
 \resizebox{\columnwidth}{!}{%
	\begin{tabular}{cccc} 
		\hline
		  Start date (UTC) &  Duration (\si{\s})  & Sample time (\si{\micro\s}) & Number of channels \\
		\hline
		2022-09-18 10:08:48 & \num{7188} & \num{51.2} & \num{512}\\ 
        2022-09-18 23:58:13 & \num{8983} & \num{51.2} & \num{512}     \\
        2022-11-19 23:10:17  & \num{4753} & \num{51.2} & \num{1024}    \\
        2023-02-25 18:53:49 & \num{1276}& \num{51.2} & \num{512}\\
        2023-06-29 14:54:23  & \num{5390}&  \num{25.6} & \num{512}\\
        2023-08-19 03:10:42 & \num{10650} & \num{25.6} &
        \num{512}\\
        2023-11-17 01:41:03  & \num{7107}& \num{25.6} & \num{512}\\
		\hline
	\end{tabular}
 }
 \tablefoot{Start time of the FRB~20121102A observations taken with the 100-m Effelsberg radio telescope in UTC, their duration in seconds, sample time and number of frequency channels. All observations have the same central frequency at \SI{1400}{\mega\hertz} and bandwidth of \SI{400}{\mega\hertz}.}
 
\end{table}

\section{Follow-up}
\label{sec:observation_details}

 The dataset consisted of \SI{12.5}{\hour} of observations of FRB~20121102A taken in September 2022, February 2023, June 2023, August 2023, and November 2023. These observations were not scheduled based on the activity of the source but instead were carried out as blind observations when the telescope was available. The data were taken with the Effelsberg 100-m telescope located in Germany. The observation details are presented in \cref{tab:example_table}. The telescope has a system equivalent flux density (SEFD) of \SI{17}{\jansky} and a minimum fluence threshold of \SI{0.15}{\jansky\milli\s} assuming bursts with a \num{1} ms duration, a \SI{300}{\mega\hertz} bandwidth, and a minimum signal-to-noise ratio (S/N) of \num{7} \citep{2021MNRAS.500..448C}. The telescope was pointed at \texttt{RA}: 05:31:58.600 and \texttt{Dec}: +33:08:49.600 to look for FRB~20121102A events. At the start of each scheduling block, we started with a short \numrange{2}{3} minute observation of the bright pulsar PSR B0355+54 (pointing to \texttt{RA}: 03:58:53.7000 and \texttt{Dec}: 54:13:13.8000). This observation was conducted to verify that the system setup and conditions for observing FRB~20121102A were optimal. 

The observations were carried out with the central beam of the seven-beam receiver (a description of this can be found in \citealt{2021MNRAS.508..300C}) in combination with Effelsberg's direct digitalisation system with a \SI{400}{\mega\hertz} bandwidth at a central frequency of \SI{1400}{\mega\hertz} with several time and frequency resolutions, going from \SI{51.2}{\micro\s} and \num{512} frequency channels to \SI{25.6}{\micro\s} and \num{1024} frequency channels in the largest dataset. 

\section{Data processing}
\label{sec:data_processing}

The data were in \texttt{PSRFITS} \citep{2004PASA...21..302H} format with four Stokes parameters and were converted to intensity \texttt{filterbanks} using the \texttt{digifil} routine from \texttt{dspsr} \citep{2011PASA...28....1V}. Once converted, the files were processed using a custom \texttt{PRESTO}-based \citep{2011ascl.soft07017R} pipeline implemented in Python. The pipeline was tested using data collected in each observing epoch from the test pulsar, PSR B0355+54. For each epoch, single pulses from this source were found down to a S/N of \num{7}, showing that the pipeline was working correctly. The pipeline executes the radio frequency interference (RFI) mitigation routine \texttt{rfifind} to generate a mask file to be used in the next stages of processing. The data of FRB~20121102A were incoherently dedispersed \citep{2004hpa..book.....L} using a DM range from \SIrange{550}{570}{\parsec\per\centi\m\cubed} with a step of \SI{1}{\parsec\per\centi\m\cubed}, and a downsampling factor of \num{8} was applied to the time series. We searched for candidates down to a minimum S/N of \num{7}. The time series for the different DM trials were used to run the single pulse search. After obtaining the pulse candidates, we removed duplicates by clustering events whose arrival time was within a window defined by the dispersive delay of the DM trials and kept those with the highest S/N. The final candidates were plotted with a custom-made waterfall plot and visually inspected to determine whether they were real events, RFI, or other artefacts.

\begin{figure*}[h]
    \sidecaption
    \centering
    \includegraphics[width=12cm]{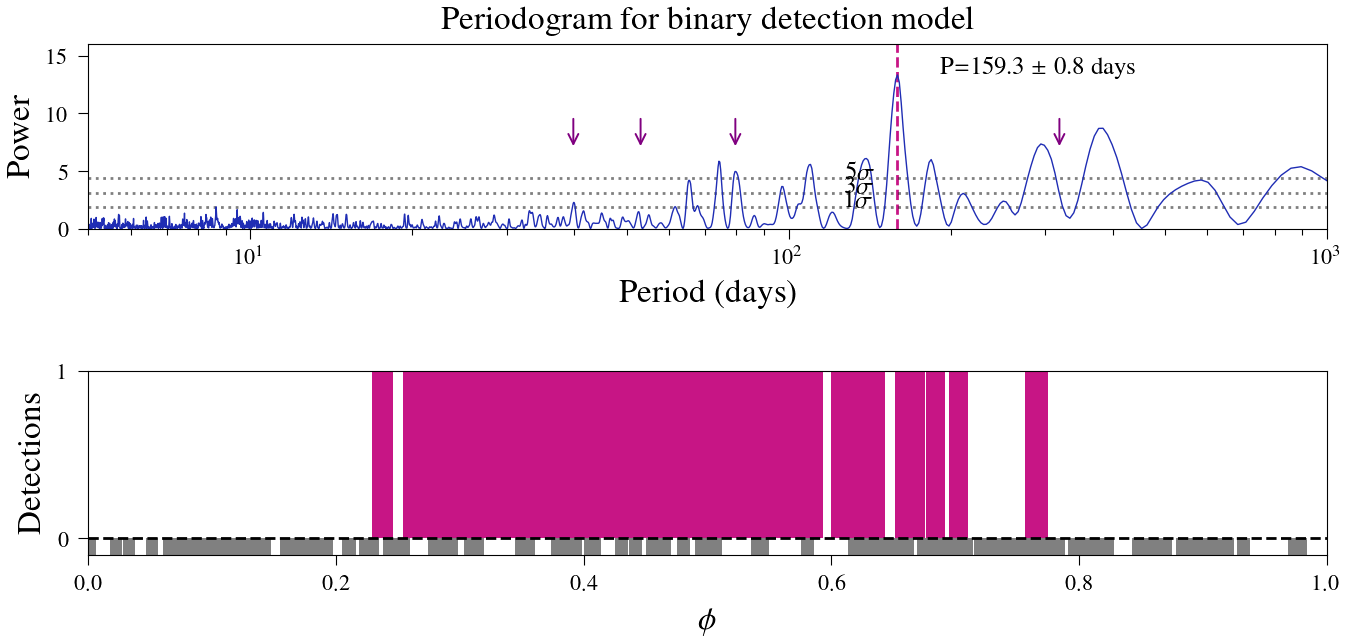}
    \caption{Periodogram for binary detection model. \textit{Top}: L-band dataset periodogram resulting from the binary model, with detection labels (0 for non-detections and 1 for detections). The first candidate of the periodogram yields a periodicity of \num{159.3+-0.8} days. The arrows indicate some of the harmonics of the period found. Two peaks to the right around 290 and 380 days are also prominent. The dotted horizontal lines correspond to the $1\sigma$, $3\sigma$, and $5\sigma$ significance levels determined by \num{10000} bootstrap resamplings. \textit{Bottom}: Folding of the observations with a periodicity of \num{159.3} days. MJD \num{58356.5} is used as the reference epoch, with phase $\phi = 0$. The detections, in magenta, are labelled 1 and the non-detections, in grey, are labelled 0 (a little height was added to these so they are visible in the plot).}
    \label{fig:binary-periodogram}
\end{figure*}
\begin{figure*}[h]
    \sidecaption
    \centering
    \includegraphics[width=12cm]{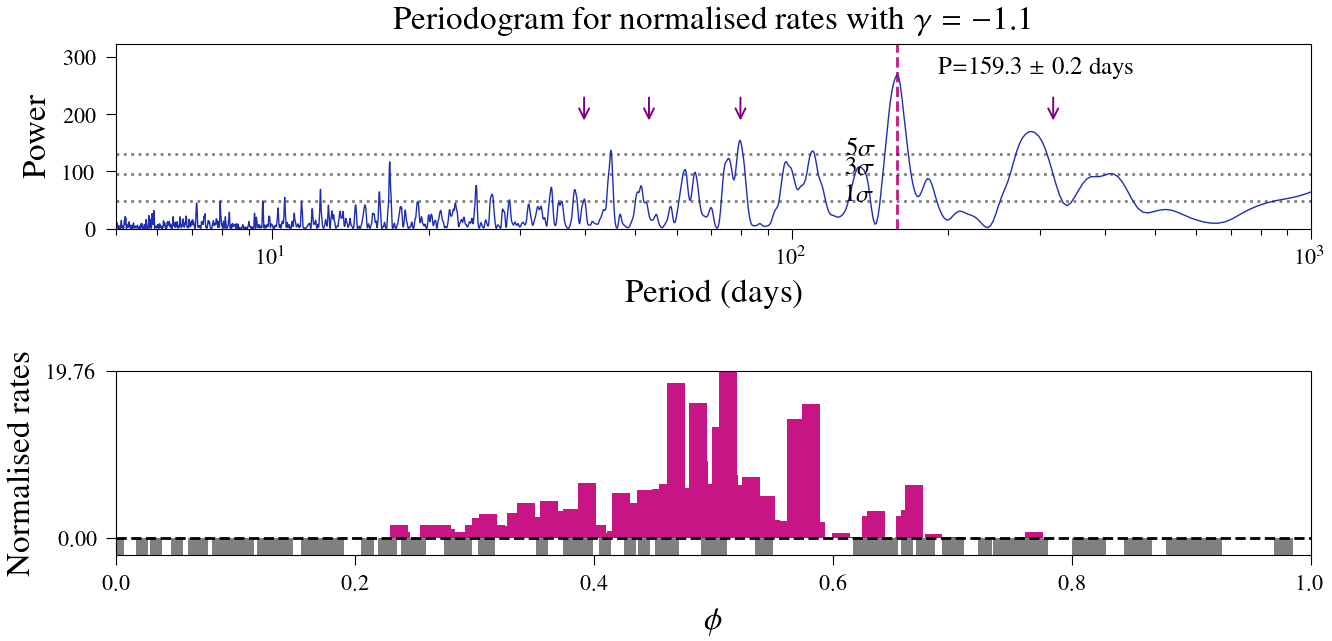}
    \caption{Periodogram for normalized rates. \textit{Top}: L-band dataset periodogram for the normalised rates model. The arrows indicate some of the harmonics of the \num{159}-day period. The peaks to the right of the plot are at roughly \num{290} and \num{380} days. They arise due to the superposition of the spacing of the two most active epochs of FRB~20121102A plus the second harmonic of the \num{159} days. The dotted horizontal lines correspond to the $1\sigma$, $3\sigma$, and $5\sigma$ significance levels determined by \num{10000} bootstrap resamplings. \textit{Bottom}: Observation-folding at \num{159} days. Observations with detections are highlighted in magenta and non-detections in grey. A little height was added to the non-detections for visualisation of phase domain coverage. The activity phase described by the magenta bars shows a Gaussian-like profile, where the detections peak in the centre.}
    \label{fig:norm-rate-periodogram}
\end{figure*}

\section{Results}
\label{sec:results_and_periodicity_calculation}

\subsection{Non-detections}
\label{subsec:nondetections}

No bursts are detected in \SI{12.5}{\hour} of data of FRB~20121102A in the L-band frequencies taken with Effelsberg, down to a S/N of 7. This means no detections down to a fluence\footnote{In this work we computed fluence based on the band-averaged peak flux density.} of \SI{0.29}{\jansky \milli\second} considering the average of the pulse width\footnote{We refer to the band-averaged pulse width as the pulse width.} from the bursts reported in \citet{2021MNRAS.500..448C}. This fluence corresponds to an isotropic energy of \SI{1.31e38}{\erg}.

\subsection{Periodicity}

\label{subsec:periodicity}
We added the observations detailed in \cref{tab:example_table} to a dataset containing several observation dates gathered from the literature. The dataset contains observations made with Effelsberg, the Arecibo Observatory with the Arecibo L-Band Feed Array (ALFA) and L-wide receivers, the Westerbork Synthesis Radio Telescope (WSRT), the Five-hundred-meter Aperture Spherical radio Telescope (FAST), the Green Bank Telescope (GBT), the Lovell Telescope, the Meerkat Telescope, the Very Large Array (VLA), and the Deep Space Network telescopes 43 and 63 (DSS-43 and DSS-63). For telescope details see \cref{fluence-table}. As there is evidence of chromaticity at the start and during the active windows of FRB~20180916B and FRB~20121102A \citep{2023MNRAS.524.3303B, 2021Natur.596..505P}, we separated the datasets of observations into L, S, and C bands, where the L band covers frequencies from \SI{1}{\giga\hertz} to just below \SI{2}{\giga\hertz}, the S band spans from \SI{2}{\giga\hertz} to just below \SI{4}{\giga\hertz}, and the C band ranges from \SI{4}{\giga\hertz} to just below \SI{8}{\giga\hertz}.

\subsubsection{L-band dataset}
The dataset is made from the observations reported in \citet{2014ApJ...790..101S}, \citet{2016ApJ...833..177S}, \citet{2016Natur.531..202S}, \citet{2017ApJ...846...80S}, \citet{2017MNRAS.472.2800H}, \citet{2017ApJ...834L...8M}, \citet{2019A&A...623A..42H}, \citet{2020A&A...635A..61O}, \citet{2020MNRAS.495.3551R}, \citet{2020MNRAS.496.4565C},
\citet{2021Natur.598..267L}, \citet{2021MNRAS.500..448C}, \citet{2022MNRAS.515.3577H}, \citet{2023MNRAS.519..666J}, \citet{2023ATel15980....1F}, \citet{2024SciBu..69.1020Z}, and the observations reported in this work. We made sure to only include each observation once, as some were reported more than once by different authors. The timestamps used correspond to the modified Julian date (MJD) marking the start of each observation. In cases where the only reported MJDs were of detected bursts, we used the MJDs of the bursts and the wait times between them to determine the approximate start of observation times. We included in the dataset a single burst reported in \citet{2023ATel15980....1F} because there is a lack of detections between \num{2021} and \num{2023}. The full dataset consisted of \num{284} epochs in a total time span of \num{4031} days, with \num{165} non-detections and \num{119} detections. We used the Lomb-Scargle periodogram \citep{2018ApJS..236...16V} technique to find the periodicity of the source and find a periodicity of \num{159.3+-.8} days, which is in very good agreement with the \num{161.3+-5} days that \citet{2021MNRAS.500..448C} find.

We tried two different methods to model the detections, in order to calculate the periodicity: a binary model and a normalised rates model. In the first model, detections are labelled \num{1} and non-detections are labelled \num{0}, in a similar way to in \citet{2021MNRAS.500..448C}. We obtain a periodicity of \num{159.3+-0.8} days for the binary model. We estimated the false alarm probability of the peak by using a bootstrap resampling with \num{10000} trials. We obtain a peak significance above $5\sigma$ and a lower uncertainty than \citet{2021MNRAS.500..448C}. \Cref{fig:binary-periodogram} shows the results of the periodogram.

The $1\sigma_\text{LS}$ uncertainty for the period was calculated using the full width at half maximum ({FWHM}) of the Gaussian fit to the peak, as formulated by \citet{2018ApJS..236...16V}:
\begin{equation}\label{eq:ls-uncertainty}
    \sigma_\text{LS} = \frac{\text{FWHM}}{2}\sqrt{\frac{2}{N \times \left(\text{S/N}\right)^{2}}},
\end{equation}
where ${N}$ is the number of points in the dataset.

As well as the binary detection model, we used a more sophisticated model
in which we instead used the rates of detection, calculated as the number of detected bursts within the observation length for each observation in our dataset. To address the difference in sensitivity across the telescopes, we translated the rates to what a 100 m telescope would infer. We refer to this rate as the normalised rate, and we calculated it as
\begin{equation}\label{eq:rates}
    R_\mathrm{norm} = R_\mathrm{ref} \left( \frac{\mathrm{F}_\mathrm{min}}{\mathrm{F}_\mathrm{ref}} \right) ^{\gamma},
\end{equation}
where $R_\mathrm{ref}$ is a reference rate obtained from an observation, $\mathrm{F}_\mathrm{min}$ is the minimum detectable fluence of the telescope we used to normalise, $\mathrm{F}_\mathrm{ref}$ is the minimum detectable fluence of the telescope from which we took the reference rate at a given observing frequency, and $\gamma$ is the power-law value that describes the cumulative energy distribution of the bursts for a given FRB. We consider a gamma value of \num{-1.1} as reported in \citet{2021MNRAS.500..448C}. The fluence of a pulse is given by (radiometer equation)
\begin{equation}\label{eq:minflux}
    F= \frac{\text{S/N} \times \mathrm{SEFD}}{\sqrt{n_{p} \times \Delta \nu}} \times \sqrt{\Delta t},
\end{equation}
 where the S/N is taken to be \num{7}, $n_{p}$ is the number of polarisations and is taken to be 2, $\Delta t$ is the pulse width and is considered to be \SI{1}{\milli\s}, and $\Delta\nu$ is the observation bandwidth. 
One bias in this model is that, although the detections are weighted relative to each other by the event rate, all non-detections are weighted equally, regardless of their duration. To mitigate this, we split the non-detections into 1-hour blocks (each representing zero events per hour). We removed the blocks of non-detections that were left with a duration of less than 1 hour after the split. We consider that the bias in this approach is less significant than treating each non-detection observation as an independent data point regardless of its duration.

The periodogram for this dataset yields a periodicity of \num{159.3+-0.2} days, which is consistent with what we obtained from the binary model. As shown in \cref{fig:norm-rate-periodogram}, the significance of this peak is greater than $5\sigma$, but lower than that from the binary model. This is expected because, in comparison with the binary model, the same number of observations are now split into more bins, given the continuous value an inferred rate can have. Other significant peaks to the left of the \num{159.3}-day period correspond to harmonics, and the two most prominent peaks to the right of \num{159.3} days are a result of the time difference between the two most active reported windows for FRB~20121102A, which are visible in \cref{fig:active-phases}.

\subsubsection{S-band dataset}
\label{subsec:S-band_dataset}

\begin{figure*}[h!]
    \sidecaption
    \centering
    \includegraphics[width=12cm]{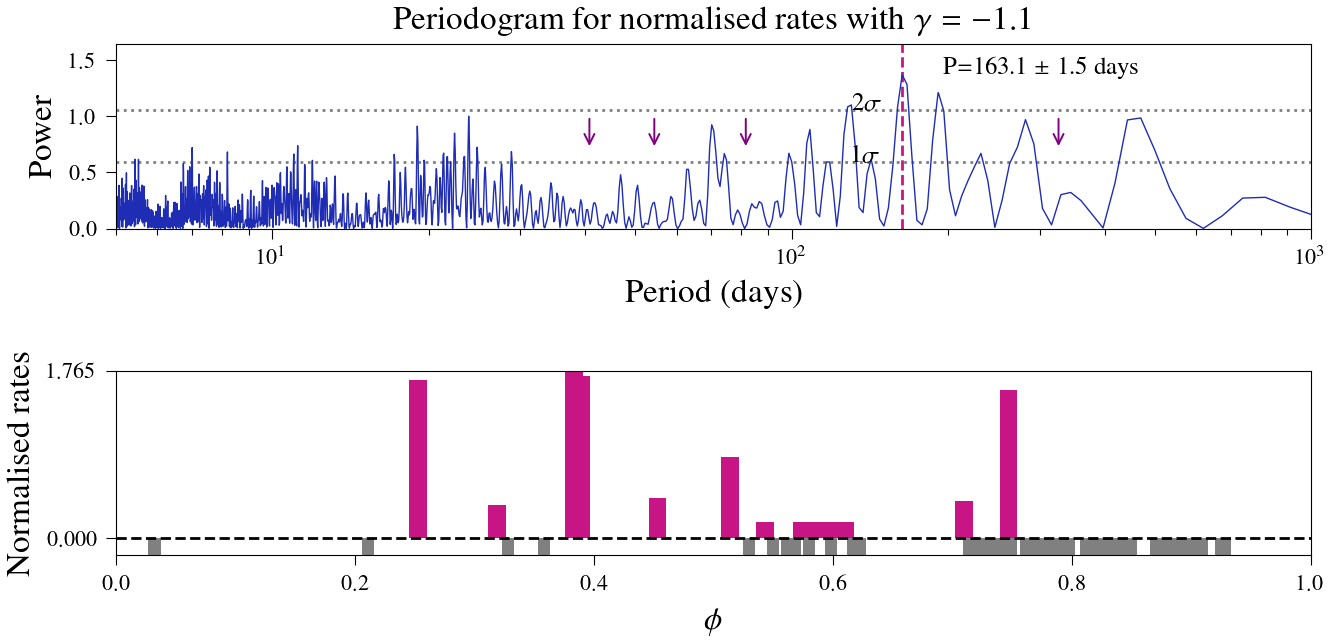}
    \caption{Periodogram for normalised rates. \textit{Top}: S-band dataset periodogram for the normalised rates model. The periodicity that this dataset yields is \num{163.1+-1.5} days. The purple arrows indicate some of the harmonics of the period. The dotted horizontal lines correspond to the $1\sigma$ and $2\sigma$ significance level determined by \num{10000} bootstrap resamplings. The peak at \num{163.1} days is slightly above $2\sigma$. \textit{Bottom}: Folded normalised rates at the obtained frequency vs phase. Observations with detections are highlighted in magenta and the non-detections in grey. A little height was added to the non-detections for visualisation of phase domain coverage. No indication for a Gaussian-like activity window is present; however, the phase domain is not fully sampled.}
    \label{fig:norm-rate-S-periodogram}
\end{figure*}

The same process was used for a dataset of observations made in the S band. The observations were taken from \citet{2016ApJ...833..177S}, \citet{2017ApJ...850...76L}, \citet{2017ApJ...846...80S}, \citet{2020ApJ...897L...4M}, \citet{2020ApJ...905L..27P}, and \citet{2021MNRAS.500..448C}. This dataset is composed of \num{80} epochs, with \num{64} non-detections and \num{16} detections for a total time span of \num{1549} days. The periodicity for this dataset is \num{159.8+-1.4} days for the binary model and \num{163.1+-1.5} days for the normalised rates model, which is in agreement with the previous two results in the L-band dataset. The binary model periodogram for this dataset is shown in \cref{fig:binary-S-periodogram} and the normalised rates model periodogram in \cref{fig:norm-rate-S-periodogram}.

\subsubsection{C-band dataset}
\label{subsec:C-band_dataset}

The dataset of observations from the C band was taken from \citet{2016ApJ...833..177S}, \citet{2017ApJ...850...76L}, \citet{2018ApJ...863..150S}, \citet{2018Natur.553..182M}, \citet{2018ApJ...863....2G}, and \citet{2021MNRAS.500..448C}. This dataset is composed of 49 epochs, with 43 non-detections and 6 detections for a total time span of \num{673} days. We find no periodicity for this dataset. This is likely due to the low number of detections present in the data.

\section{Discussion}
\label{sec:discussion}

\begin{figure*}[h]
    \sidecaption
    \centering
    \includegraphics[width=12cm]{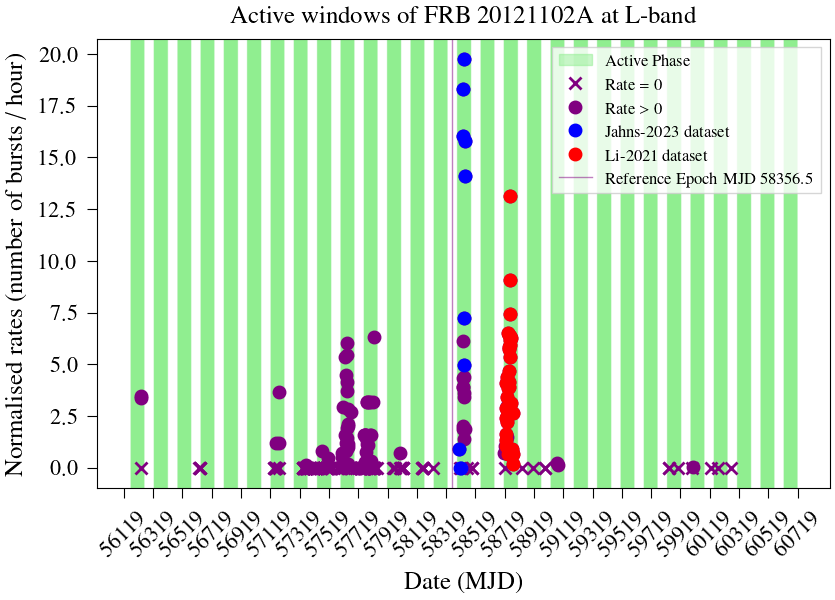}
    \caption{Normalised rates vs MJD, spanning from the oldest window where FRB~20121102A has been observed in the L band to the current active window. The active phase of FRB~20121102A, displayed with the green bands, corresponds to \SI{53}{\percent} of the phase and a periodicity of \num{159.3 +- 0.8} days. The reference epoch is MJD \num{58356.5} for a phase of \num{0}. The observations from November rain \citep{2023MNRAS.519..666J} and FAST's sample \citep{2021Natur.598..267L} are highlighted in blue and red, respectively.
    }
    \label{fig:active-phases}
\end{figure*}
The results obtained for the periodicity of FRB~20121102A are consistent and are a further improvement over those previously reported in \citet{2021MNRAS.500..448C} because our dataset contains more epochs and an extended time span. We tried the method of normalising the rates to test whether that improved the estimation of the periodicity of FRB~20121102A and find a periodicity of \num{159.3 +- 0.2} days. The advantage of the normalised rates model is that it weighs each timestamp by the level of activity within it. In addition, the binary model has \num{159.3 +- 0.8} days, and both peaks have a significance greater than $5\sigma$, but the normalised rates model has a lower significance than the binary model. We attribute the difference to the fact that, in the binary model, the power of the signal is split between detections (1) and non-detections (0), while in the normalised rates model the power is distributed across a wider bin range. This also introduces more noise to the time series. An in-depth analysis of the different methods for estimating the periodicity is work in progress.

Additionally, a more accurate model for the normalised rates would consider different $\gamma$ values in \cref{eq:rates} for each telescope in the dataset. To test the importance of the parameter in our results, in addition to the $\gamma=\num{-1.1}$ that \cite{2021MNRAS.500..448C} obtain, we tried using $\gamma=\num{-1.37}$ as reported in \cite{2021Natur.598..267L} and obtain a consistent periodicity. Therefore, because we used Effelsberg as our reference in the normalisation, we consider that using $\gamma=\num{-1.1}$ for all telescopes is optimal. Because of the higher significance of the peak in the binary model, we take this as the new best periodicity for FRB~20121102A. To centre the activity window around a phase of $\phi=0.5$, we used MJD \num{58356.5} as a reference epoch for phase $\phi = 0$ and we obtain an activity window of \SI{53}{\percent}. This broadly means that during the \num{159} days of one cycle, the source is active within an \num{84}-day window in the L band. This is broader than the \num{31}\% that \citep{2020Natur.582..351C} find for FRB20180916B.

As \cref{fig:norm-rate-periodogram} shows, the highest peaks in the periodogram of the normalised rates are at \num{159} days and \num{290} days. A third peak at around \num{380} days is also significant. As marked with purple arrows in the periodogram, the peak at \num{290} days superposes with the second harmonic of the 159-day peak. We also see these peaks in the periodogram of the binary model, but with significantly lower power. By plotting all the observing epochs against their normalised rates as shown in \cref{fig:active-phases}, we observe that the two most active windows occurred in 2018, with the observations of Arecibo from the `November rain' \citep{2023MNRAS.519..666J}, and 2019, with the {FAST} telescope observations \citep{2021Natur.598..267L}. This introduces an artefact in the calculation of the periodicity. We can explain the peaks at 290 and 380 days by the spacing between these high-activity episodes, with intervals ranging from \num{270} days for the closest observations to \num{370} days for the furthest apart. To verify that such peaks correspond to artefacts, we folded the epochs using the above-mentioned reference epoch. The result from folding at 290 and 380 days is a distribution of epochs with detections all throughout the phase space, with no clear trend. We only find a dispersion-optimised distribution of the detection epochs for the 159-day period, as shown at the bottom of the periodogram in Fig. \ref{fig:norm-rate-periodogram}. Observations triggered by known activity of the source may introduce bias in the calculation of the periodicity. In our case, since we have \SI{58}{\percent} non-detections and \SI{42}{\percent} detections, this bias is minimised.

From the folding, we clearly see that the active phase has a Gaussian-like profile, with the peak of the detections located in the middle of the window. This behaviour is also seen for FRB~20180916B \citep{2020Natur.582..351C}. Overall, we conclude that, although the normalised rates model has greater physical meaning, more data are needed to achieve the same or higher levels of significance compared to the binary model.

\begin{figure*}[h]
    \sidecaption
    \centering
    \includegraphics[width=12cm]{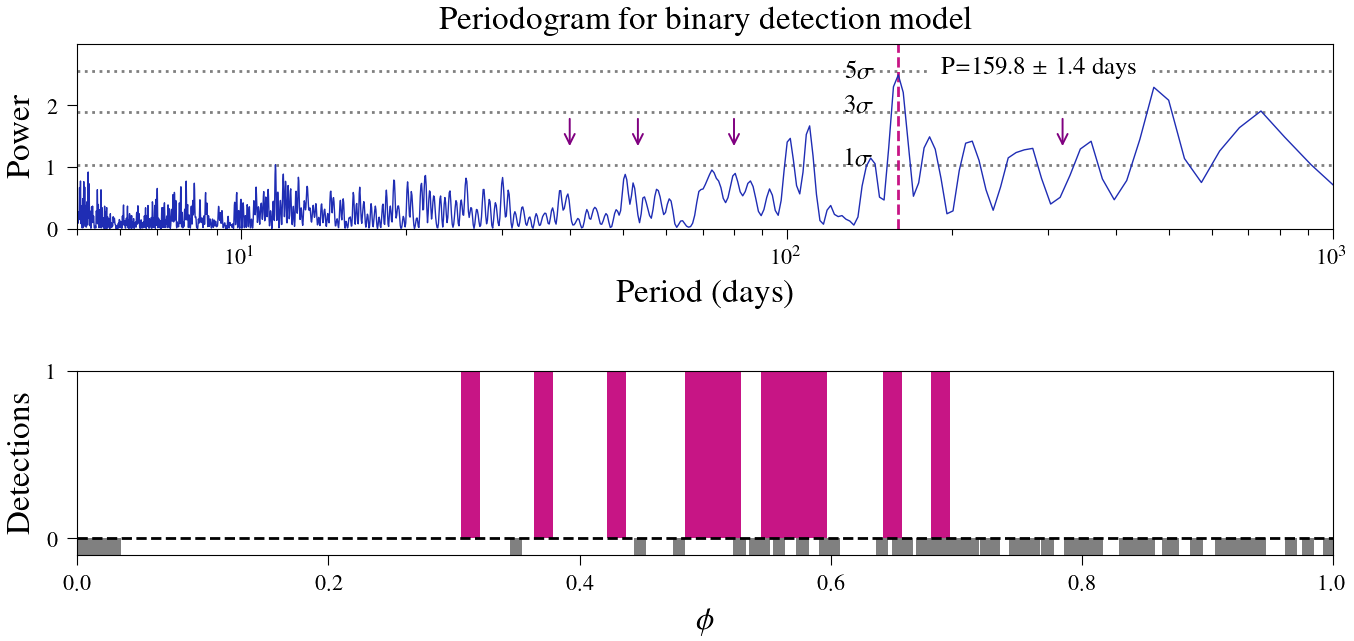}
    \caption{Periodogram for binary detection model. \textit{Top}: S-band dataset periodogram resulting from the binary model. The periodicity this dataset yields is \num{159.8+-1.4} days. The purple arrows indicate some of the harmonics of the period found. The dotted horizontal lines correspond to the $1\sigma$, $3\sigma$, and $5\sigma$ significance levels determined by \num{10000} bootstrap resamplings. The peak at \num{159.8} days is right below a $5\sigma$ significance. \textit{Bottom}: Folding of the observations using the obtained periodicity. In this dataset, more observations are needed to be able to map the full phase domain. Observations with detections are highlighted in magenta and the non-detections in grey. A little height was added to the non-detections for visualisation of phase domain coverage.}
    \label{fig:binary-S-periodogram}
\end{figure*}

Regarding the dependence of the activity window on the observing frequency, we obtain a consistent period for the L and S bands. For the normalised rate model, the 163.1-day value we obtain for the periodicity is consistent with both the S-band binary model and with the L-band value within $2.5\sigma$. This means that the phenomenon leading to the active phase of the source is intrinsic. However, as seen in the folding of the S-band dataset shown in Fig. \ref{fig:binary-S-periodogram}, the full phase range is not mapped. In particular, we lack observations in a considerable part of the phase, towards the start of the activity. We do not see a clear Gaussian-like trend like we do for the L-band dataset, probably because of the limited dataset. More observations in this band will permit a quantitative comparison across frequencies.

Folding the S-band dataset with the periodicity obtained in the L band, we find that the activity window for the S band ends before the L-band activity window, with a phase difference of 0.12. This means that the S-band activity phase finishes 19 days before the L-band active phase. Since we do not have the \numrange{0}{0.4} part of the phase completely mapped in the S band we cannot conclude with certainty what happens in this part.

Regarding the C-band dataset, we do not have enough observations, and in particular detections for the Lomb-Scargle periodogram, to obtain a consistent periodicity. More observations will help us better understand the shift in phase between frequency bands and the behaviour of the activity phase (Espinoza-Dupouy et al. in prep.).

Using our \num{159.3}-day periodicity, with a 53\% active phase, and our reference epoch of \num{58356.5}, we calculate that as of late November 2024 the source was at 0.36 phase and that the next active phase should begin on 19 April 2025 UTC and end on 12 July 2025 UTC. We emphasise that this does not guarantee detections of the source, but rather that it indicates when the source is most likely to emit.

Our result for the periodicity is consistent with the recent work of \cite{2024ApJ...969...23L}. Unlike their approach, which considers the MJDs of the bursts independently of the frequency band, we separated the observing epochs based on the frequency at which the observations were conducted. \cite{2024ApJ...969...23L} report a candidate period of $157.1^{+5.2}_{-4.8}$ and another of $4.605^{+0.003}_{-0.010}$ days, using a phase-folding algorithm. The second candidate period was not detected in their Lomb-Scargle periodogram, nor does it appear in ours. Furthermore, we folded the dataset at the trial period and do not see any trend. Given the chromatic behaviour that \citep{2023MNRAS.524.3303B} observe in the active windows of FRB~20180916B and the hint of dependence in FRB~20121102A presented in our work, we believe a frequency separation to be necessary. Naturally, more observations across the frequency range will help pinpoint the magnitude of the shift and the duration of the activity window.

The widely accepted scenarios to explain periodic actively repeating FRBs involve either a precessing magnetar or a neutron star (NS) in a binary system. In the literature, several precession models of hot, young, non-superconducting, and highly active long-period isolated magnetars can explain the periodicity of FRB~20180916B \citep{2020MNRAS.496.3390B, 2024MNRAS.530.3641F}. However, most of these models do not consider vortex superfluidity, which can dampen long-period precession \citep{1977ApJ...214..251S}, therefore making this explanation for the periodicity of FRB~20121102A unlikely. Although there have been attempts to reconcile long-period precession with quantum vortices, reconciliation seems to occur only under very specific conditions, such as with a weakly coupled toroidal magnetic field \citep{2019MNRAS.482.3032G}.

Alternatively, models of NSs in binary systems have been proposed to explain the longer timescale periodicity of FRB~20121102A \citep{2020ApJ...893L..39L, 2020ApJ...893L..26I}.
An NS with a black hole companion could induce spin precession on timescales similar to those observed in actively repeating FRBs. However, for this effect to occur, the NS would need to be in a close orbit around the black hole, where the typical orbital lifespan of the system is approximately 10 years. As the NS moves closer to the black hole, relativistic effects, such as precession and spin-up of the NS, should become increasingly pronounced \citep{2020ApJ...893L..31Y}. Despite this, FRB~20121102A has been observed for over a decade and its activity window has not shown the rate of change predicted by these models.

Another possible companion for an NS is a massive star that continuously emits stellar winds. As the NS interacts with the stellar wind of its companion, it can create activity windows that vary with frequency. To explain the observed DM of FRB~20121102A using these models, the FRB would need to be emitted when the NS is near the periastron of its orbit. This scenario would result in a variation of four orders of magnitude in the DM within approximately 10\% of the orbital phase \citep{2020ApJ...893L..39L}. However, observations indicate that the activity window of FRB~20121102A spans about half of its cyclic period, and no DM variation of that magnitude has been observed.

\section{Conclusions}
\label{sec:conclusions}
We report seven observing epochs of FRB~20121102A monitoring and find no detections. We combined our observations with publicly available data on the source, dividing this combined dataset by frequency band, and find a new best periodicity for the cyclic activity window to be \num{159.3 +- 0.8} days at L-band. This updated periodicity is more precise and has a higher significance than that previously reported in \cite{2021MNRAS.500..448C}, mainly because of the larger and extended dataset. We also report the same periodicity using a normalised rates model that is more precise but has a lower significance than the one obtained by the binary model. We find a consistent periodicity in the S band and report that this active phase seems to end 19 days before the L-band active phase. No conclusions can be drawn from the C-band observations given the lack of observing epochs and, in particular, of detections. We strongly encourage the community to report both detections and non-detections of repeating FRBs, along with full observation details. Sky and instrumentation statistics such as exposure time, observation duration, and telescope properties are fundamental to further constraining the nature of FRBs.

\begin{acknowledgements}
This publication is based on observations with the 100-m telescope of the Max-Planck-Institut f\"{u}r Radioastronomie at Effelsberg. C.B would like to thank the Max Planck Partner group at PUC led by M. C. for funding the internship that led to this work. C. B. would like to thank B. Briceño for his support and feedback during this project. T. C. acknowledges support by the ANID BASAL FB210003 and fondo de astronomía: ANID / Fondo 2023 QUIMAL/ QUIMAL230001. 
\end{acknowledgements}

\bibliographystyle{aa}
\bibliography{aanda.bib}
\newpage

\newpage
\begin{table}[h]
\centering
\caption{Sensitivity of telescopes in the sample.}
\label{fluence-table}
\begin{tabular}{cccc}
\hline
Telescope & $F_{min}$ (Jy ms) & Frequency Band \\ \hline
Effelsberg          & 0.15             & L-band          \\ 
Arecibo             & 0.03 / 0.02      & L-band (ALFA / L-Wide) \\ 
GBT                 & 0.05             & S-band          \\ 
FAST                & 0.012            & L-band          \\ 
WSRT                & 0.26             & L-band          \\ 
Meerkat             & 0.06             & L-band          \\ 
DSS-43              & 0.23             & S-band          \\ 
DSS-63              & 0.5              & S-band          \\ 
VLA                 & 0.19 / 0.06      & L-band / S-band \\ 
Lovell              & 0.23             & L-band          \\ \hline
\end{tabular}
\tablefoot{The first column displays the names of the radio telescopes, the second the minimum fluence values ($F_{min}$) for each telescope and the third column indicates the specific frequency band used.}
\end{table}

\end{document}